\documentclass[aps,pra,superscriptaddress,twocolumn]{revtex4-1}

\usepackage{graphicx} 
\usepackage{amsmath}
 
\newcommand{\beq}{\begin{equation}}
\newcommand{\enq}{\end{equation}}
\newcommand{\bea}{\begin{eqnarray}}
\newcommand{\ena}{\end{eqnarray}}

\newcommand{\ad}{\hat{a}^\dagger}
\newcommand{\ha}{\hat{a}}
\newcommand{\rr}{{\bf r}}

\begin{document}

\title{Kelvin-Helmholtz instability in two-component Bose gases on a lattice}
\author{E. Lundh}
\affiliation{Department of Physics, Ume{\aa} University, 901 87 Ume\aa, Sweden}
\author{J.-P. Martikainen}
\affiliation{Nordita, 106 91 Stockholm, Sweden}
\affiliation{Aalto University, P.O. Box 15100, FI-00076 Aalto, Finland}
\date{\today}
\begin{abstract}
We explore the stability of the interface between two phase-separated 
Bose gases in relative motion on a lattice. 
Gross-Pitaevskii-Bogoliubov theory and the Gutzwiller ansatz
are employed to study the short- and long-time stability properties. 
The underlying lattice introduces effects of discreteness, broken spatial 
symmetry, and strong correlations, all three of which are seen to 
have considerable qualitative effects on the Kelvin-Helmholtz instability. 
Discreteness is found to stabilize low flow velocities, because of the 
finite energy associated with displacing the interface. 
Broken spatial symmetry introduces a dependence not only on the relative 
flow velocity, but on the absolute velocities. 
Strong correlations close to a Mott transition will 
stop the Kelvin-Helmholtz instability from affecting the bulk density 
and creating turbulence; instead, the instability will excite vortices with 
Mott-insulator filled cores. 

\end{abstract}
\maketitle

\section{Introduction}
\label{sec:intro}
The Kelvin-Helmholtz instability is a dynamical instability of the interface
between two fluids that move relative to one another. This 
hydrodynamical instability 
can occur in very different settings, for example when
the wind is blowing over the surface of the smooth ocean or 
in plasma flows in the Earth's magnetic field~\cite{Hasegawa2004a}, 
or at the $AB$-phase boundary in 
superfluid $^{3}{\rm He}$~\cite{Blaauwgeers2002a,Volovik2002a}. 
The creation of degenerate mixtures of weakly interacting different bosonic 
gases~\cite{Hall1998b,Stenger1998a,Modugno2001a} enables
studies of the Kelvin-Helmholtz instability in quantum systems
which are more controllable as well as easier to measure in detail
than  more strongly interacting 
superfluids such as liquid $^{3}{\rm He}$.
Takeuchi, Suzuki {\it et al.}~\cite{takeuchi2010,suzuki2010} performed 
theoretical studies of 
the Kelvin-Helmholtz instability in a two-component condensate. 
Essentially, Ref.~\cite{takeuchi2010} confirmed the expectation that 
Bose-Einstein 
condensed atomic gases provide an ideal setting to study the classic 
Kelvin-Helmholtz instability without viscosity complicating the picture. 
Ref.\ ~\cite{suzuki2010} found a considerably altered instability 
dispersion relation in the case of a wide interface with large 
overlap between the two condensates; this was classified  
as a counter-superflow instability.

Putting a Bose gas in an optical lattice introduces several new features. 
Most dramatically, the gas exhibits a quantum phase transition 
between a superfluid and a Mott insulating state when the ratio 
between the tunneling and interaction energies passes a 
critical value \cite{jaksch1998}. In a binary condensate the phase 
diagram is richer, displaying different combinations of Mott insulator 
and superfluid in coexisting or phase separated configurations 
\cite{powell2009,iskin2010}. Moreover, even in the superfluid state where a 
significant portion of the atoms are Bose-Einstein condensed, the discreteness 
and the broken translational symmetry can have decisive effect on the 
motion of the bosons, as we shall see. 

In this paper, we study the Kelvin-Helmholtz-type instabilities of a 
phase separated binary condensate on a lattice. In Sec.\ \ref{sec:model}, 
we present the equations that we work with. In Sec.\ \ref{sec:weak}, 
we study a weakly segregated pair of condensates in order to explore 
the effects of broken translational symmetry on the interface dynamics. 
In Sec.\ \ref{sec:strong}, we study instead a strongly segregated 
pair of condensates to see the effects of discreteness of the lattice. 
Sec.\ \ref{sec:analytic} presents a heuristic derivation of a 
closed-form expression for the numerically obtained instabilities. 
In Sec.\ \ref{sec:mott}, we study the system close to a Mott 
transition in order to explore the effects of strong correlations. 
Finally, in Sec.\ \ref{sec:conclusion}, we summarize and conclude.

\section{Theory of a two-component lattice Bose gas}
\label{sec:model}
The system is assumed to consist of two species of bosons at 
nearly zero temperature hopping on 
a square lattice, as can be realized with evaporatively cooled atoms in 
an optical lattice \cite{Hall1998b,Stenger1998a,Modugno2001a}. 
If the lattice is deep enough, 
the tight-binding approximation is valid and the bosons can be  
assumed to occupy the lowest band only. 
For simplicity, we assume the two species to have equal masses, 
interaction constant and tunneling properties; this can be 
realized by choosing the two components to be two spin states of 
the same element. 
The many-body Hamiltonian governing this system is 
\bea
\hat{H} &=& 
- \sum_{j=1}^2\sum_{<rr'>} J_j\ad_{jr} \ha_{jr'}\nonumber\\
&+& \frac{1}{2} \sum_{j=1}^2 U_{jj}\sum_{r}\ad_{jr}\ad_{jr} \ha_{jr} \ha_{jr} 
\nonumber\\
&+& U_{12} \sum_{r} \ad_{1r} \ha_{1r} \ad_{2r} \ha_{2r} 
- \sum_{j=1}^2 \mu_j\sum_r \ad_{jr} \ha_{jr}. 
\label{hamiltonian}
\ena
Here, $J_j$ are the tunneling matrix elements, $U_{jj}$ are the in-species 
interaction parameters and $U_{12}$ is the inter-species interaction 
parameter. $\mu_j$ is the chemical potential of species $j$. 
The summation index $j=1,2$ denotes the different species 
and the index $r$ runs over the lattice sites. We will be studying 
a two-dimensional square lattice in this paper, and we therefore pass 
to a vector notation, ${\bf r}=(x,z)$, where we let the 
dimensionless Cartesian coordinates $x$ and $z$ take on 
integer values. 

For strong enough hopping and for high enough density, the lattice gases are
almost entirely Bose-Einstein condensed and the problem can be 
treated using Gross-Pitaevskii and Bogoliubov analysis. 
This regime is suitable for identifying effects of the discreteness 
and broken translational invariance of the lattice, effects that 
do not depend on quantum fluctuations. 
In this regime, the lattice gas 
is accurately described using 
condensate wavefunctions $\Phi_{j}({\bf r},t) = \langle \ha_{jr}\rangle$,  
whose dynamics follows from the discrete 
two-component Gross-Pitaevskii (GP) equation
\begin{widetext}
\begin{eqnarray}
	i\frac{\partial\Phi_{1}({\bf r},t)}{\partial t}&=&-J_1
\nabla^2\Phi_1({\bf r},t)+\left[U_{11}|\Phi_1|^2+U_{12}|\Phi_2|^2\right]
\Phi_1({\bf r},t),\nonumber\\
i\frac{\partial\Phi_{2}({\bf r},t)}{\partial t}&=&-J_2
\nabla^2\Phi_2({\bf r},t)+\left[U_{22}|\Phi_2|^2+U_{12}|\Phi_1|^2\right]
\Phi_2({\bf r},t),
\end{eqnarray}
\end{widetext}
where the discrete Laplacian is defined as 
\beq
\nabla^2\Phi({\bf r}) = \sum_{{\bf r}'}\Phi({\bf r'}),
\label{discretelaplacian}
\enq
and the sum over ${\bf r}'$ runs over the nearest neighbors to the site $\rr$.

\subsection{Bogoliubov approach}

In the Bogoliubov approximation one assumes 
for each component a stationary wavefunction with a 
small time-dependent perturbation,
\bea
	\Phi_1({\bf r},t)=\left[\Psi_1(z,t)+\delta\Psi_1(x,z,t)\right]
	\exp(-i\mu t),
\nonumber\\
	\Phi_2({\bf r},t)=\left[\Psi_2(z,t)+
	\delta\Psi_2(x,z,t)\right]\exp(-i\mu t).
\ena

The equation of motion to zeroth order is the time-independent 
1D GP equation that 
is used to calculate the interface profile,
\begin{widetext}
\begin{eqnarray}
-J_1\frac{\partial^2\Psi_1(z)}{\partial z^2}+
\left[U_{11}|\Psi_1(z)|^2+U_{12}|\Psi_2(z)|^2\right]
\Psi_1(z)
&=&
\mu_1\Psi_{1}(z), \nonumber\\
-J_2\frac{\partial^2\Psi_2(z)}{\partial z^2}+
\left[U_{22}|\Psi_2(z)|^2+U_{12}|\Psi_1(z)|^2\right]
\Psi_2(z)
&=&
\mu_2\Psi_{2}(z),
\end{eqnarray}
\end{widetext}
where we introduced notation analogous to Eq.\ (\ref{discretelaplacian}),
\beq
\frac{\partial^2\Psi_j(z)}{\partial z^2} = \Psi_j(z+1)+\Psi_j(z-1).
\enq

Anticipating that the nonlinear term in the GP equation will couple 
positive- and negative-frequency modes, we define in a standard way 
\beq
	\delta\Psi_i({\bf r},t)=\left[u_i({\bf r},t)\exp(-i\omega t)+
	v_i^*({\bf r},t)\exp(i\omega t)\right]\exp(-i\mu t).\nonumber
\enq

Ignoring the second order terms in the perturbations we get
the Bogoliubov equations for the discrete 
double condensate system,
\beq
	{\hat M}{\bf v}=\omega{\hat \eta}{\bf v},
\enq
where
\beq
 {\hat M}=\left(\begin{array}{cccc}
H_{1} & U\Psi_1^2 & C & D\\
U\Psi_1^{*2} & H_{1} & D^{*} & C^{*}\\
C^* & D & H_{2} & U\Psi_2^{2}\\
D^{*} & C & U\Psi_2^{*2} & H_{2}
\end{array}\right),
\label{bogoliubovmatrix}
\enq

\beq
 {\hat \eta}=\left(\begin{array}{cccc}
1 & 0 & 0 & 0\\
0 & -1 & 0 & 0\\
0 & 0 & 1  & 0\\
0 & 0 & 0  & -1
\end{array}\right),  
\enq
and
\beq
{\bf v}=\left(\begin{array}{c}
u_1({\bf r})\\
v_1({\bf r})\\u_2({\bf r})\\v_2({\bf r})
\end{array}\right).
\enq
The coefficients are defined through
$C=U_{12}\Psi_1\Psi_2^*$, $D=U_{12}\Psi_1\Psi_2$, and $n_i=|\Psi_i|^2$,
\beq
	H_1=-J_1\nabla^2+2U_{11}n_1+U_{12}n_2-\mu_1,
\enq
and
\beq
	H_2=-J_2\nabla^2+2U_{22}n_2+U_{12}n_1-\mu_2.
\enq

Let us repeat some common terminology and standard results 
\cite{fetter1972}: 
The norm of a Bogoliubov mode is defined as 
\beq
||{\bf v}|| = \sum_r {\bf v}^{\dag}({\bf r})\hat\eta{\bf v}({\bf r}).
\enq 
Unless the norm is zero, we are free to rescale the Bogoliubov vector so 
that its absolute value is unity. A mode with $||{\bf v}||=1$ is 
called a quasiparticle mode and a mode with $||{\bf v}||=-1$
is called a quasihole mode. If $\Psi$ is the ground state, then 
all the quasiparticle modes have positive eigenfrequencies and 
all quasihole modes have negative eigenfrequencies; otherwise, 
no such rule applies. A mode with complex eigenfrequency has 
zero norm. 
Such modes will grow exponentially in time and thus 
they signal a dynamical instability. 
We will therefore be 
interested in the imaginary parts of the Bogoliubov eigenfrequencies in 
this article. 

The symmetries of the problem give rise to degeneracies in the 
Bogoliubov spectrum. 
First, the well-known quasiparticle-quasihole symmetry of the 
Bogoliubov equations \cite{fetter1972} carries over to the two-component case: 
Each real eigenvalue $\omega$
with an eigenvector ${\bf v}=\left(u_1(\rr), v_1(\rr),u_2(\rr), v_2(\rr)\right)^T$
has a corresponding  solution with an eigenvalue $-\omega$
and an eigenvector 
\beq
{\bf v}=-\left(v_1(\rr)^*, u_1(\rr)^*,v_2(\rr)^*, u_2(\rr)^*\right)^T.
\label{ph_symmetry_general}
\enq
In addition, imaginary eigenvalues appear in pairs of $\omega$ and $\omega^*$.
Since we will be mostly interested in a symmetric interface 
where $\Psi_1({\bf r})=\Psi_2(-{\bf r})$, we have an additional 
degenerate solution
\beq
\left(u_2(-\rr),v_2(-\rr),u_1(-\rr),v_1(-\rr)\right)^T,
\label{exch_symmetry_general}
\enq
with eigenvalue $\omega$. Combining Eqs.\ (\ref{ph_symmetry_general}-\ref{exch_symmetry_general}) to obtain 
\beq
\left(v_2^*(-\rr),u_2^*(-\rr),v_1^*(-\rr),u_1^*(-\rr)\right)^T,
\label{final_symmetry_general}
\enq
with eigenvalue $-\omega$. 
The latter two symmetries hold strictly only when equal and opposite 
currents are induced in the two components, as we will discuss 
further below.

\subsubsection{One moving condensate}
The discrete system is not translationally 
invariant, and we can therefore not use the relative velocity of the 
condensates as a unique parameter; also the absolute individual 
currents are physically relevant. In particular, we will study two 
different cases: the case where only one of the condensates carries 
a current, and that where the condensates support equal but 
counter-flowing currents. 

We first assume that component $2$ is moving along the $x$ direction and that
all other position dependence of the stationary solutions is
along $z$. Furthermore, the stationary wavefunctions can be taken as real.
This means $\Psi_1({\bf r})=\Psi_1(z)$ and
$\Psi_2({\bf r})=\Psi_2(z)\exp(ikx)$, with $-\pi<k\leq\pi$. We then have
$C=U_{12}\Psi_1(z)\Psi_2(z)\exp(-ikx)$, $D=U_{12}\Psi_1(z)\Psi_2(z)\exp(ikx)$,
and
\beq
	H_2=-J_2\nabla^2+2U_{22} n_2+U_{12}n_1-\mu_2-2J_2(1-\cos k),
\enq
where $\mu$ is the chemical potential of the second component in the 
absence of a current; the term $2J(1-\cos k)$ due to the current 
is made explicit for convenience. 
The excitations can be expanded as plane waves along the $x$ direction. 
Dependency on the $x$ coordinate is eliminated 
with the choice
\begin{eqnarray}
u_1({\bf r})= u_1(z)\exp(ik_{e}x-ikx/2),\nonumber\\
v_1({\bf r})= v_1(z)\exp(ik_{e}x-ikx/2),\nonumber\\
u_2({\bf r})= u_2(z)\exp(ik_{e}x+ikx/2),\nonumber\\
v_2({\bf r})= v_2(z)\exp(ik_{e}x-3ikx/2).
\end{eqnarray}
The Bogoliubov equations now reduce to a 1D problem, and the Bogoliubov 
matrix takes the form 
\beq
 {\hat M}_{\rm 1D}=\left(\begin{array}{cccc}
H_{1u} & U\Psi_1^2 & C & D\\
U\Psi_1^{*2} & H_{1v} & D^{*} & C^{*}\\
C^* & D & H_{2u} & U\Psi_2^{2}\\
D^{*} & C & U\Psi_2^{*2} & H_{2v}
\end{array}\right),
\label{bdg_1d}
\enq
where the diagonal elements are 
\begin{widetext}
\bea
	H_{1u}=H_{1v}=-J_1\frac{\partial^2}{\partial z^2}+2J_1
\left[1-\cos(k_e-k/2)\right]+
2U_{11}n_1+U_{12}n_2-\mu_1,
\nonumber\\
	H_{2,u_2}=-J_2\frac{\partial^2}{\partial z^2}
-2J_2(1-\cos k)
+2J_2\left[1-\cos(k_e+k/2)\right]+
2U_{22}n_2+U_{12}n_1-\mu_2,
\nonumber\\
	H_{2,v_2}=-J_2\frac{\partial^2}{\partial z^2}+
-2J_2(1-\cos k)
+2J_2\left[1-\cos(k_e-3k/2)\right)]+
2U_{22}n_2+U_{12}n_1-\mu_2.
\label{H2v2}
\ena
\end{widetext}

\subsubsection{Counter-directed currents}
In the case of equal and opposite currents in the respective 
components, we write
$\Psi_1(x,z)=\Psi_1(z)e^{ikx/2}$
and
$\Psi_2(x,z)=\Psi_2(z)e^{-ikx/2}$.
The excitations can now be written
\begin{eqnarray}
u_1({\bf r})&=& u_1(z)e^{ik_ex+ikx/2},\nonumber\\
v_1({\bf r})&=& v_1(z)e^{ik_ex-ikx/2},\nonumber\\
u_2({\bf r})&=& u_2(z)e^{ik_ex-ikx/2},\nonumber\\
v_2({\bf r})&=& v_2(z)e^{ik_ex+ikx/2}.
\end{eqnarray}
We find that the off-diagonal terms of the matrix $\hat{M}_{\rm 1D}$ 
are the same as before, but to the diagonal we get
\begin{widetext}
\bea
	H_{1,u_1}=-J_1\frac{\partial^2}{\partial z^2}
-2J_1(1-\cos k/2)+2J_1\left[1-\cos(k_e+k/2)\right]+
2U_{11}n_1+U_{12}n_2-\mu_1,
\nonumber\\
	H_{1,v_1}=-J_1\frac{\partial^2}{\partial z^2}
-2J_1(1-\cos k/2)+2J_1\left[1-\cos(k_e-k/2)\right]+
2U_{11}n_1+U_{12}n_2-\mu_1,
\nonumber\\
	H_{2,u_2}=-J_2\frac{\partial^2}{\partial z^2}
-2J_2(1-\cos k/2)+2J_2\left[1-\cos(k_e-k/2)\right]+
2U_{22}n_2+U_{12}n_1-\mu_2,
\nonumber\\
	H_{2,v_2}=-J_2\frac{\partial^2}{\partial z^2}
-2J_2(1-\cos k/2)+2J_2\left[1-\cos(k_e+k/2)\right]+
2U_{22}n_2+U_{12}n_1-\mu_2.
\ena
\end{widetext}

Expressed in the reduced 1D Bogoliubov amplitudes $u_j(z), v_j(z)$, 
the symmetries (\ref{ph_symmetry_general}-\ref{final_symmetry_general}) acquire 
new forms, since the effective Hamiltonians appearing in the diagonal
of the Bogoliubov eigenvalue problem are different.
The symmetry of greatest interest to us is that for a given $k$ and $k_e$, each real eigenvalue $\omega$
with a reduced eigenvector $\left(u_1(z), v_1(z),u_2(z), v_2(z)\right)^T$
now has a corresponding hole mode with eigenvalue $-\omega$
and reduced eigenvector 
\beq
\left(v_2(-z), u_2(-z),v_1(-z), u_1(-z)\right)^T.
\label{symmetry}
\enq

\section{Kelvin-Helmholtz instability in a discrete system}

After assuming that the two condensates have equal chemical potentials
$\mu_1=\mu_2=\mu$, 
tunneling $J_1=J_2=J$, 
and interaction strengths $U_{11}=U_{22}=U$, the system can be characterized by 
three parameters. 
These are chosen to be the ratios $J/U$ and $U_{12}/U$, and the 
number density per site far from the interface, $\bar{n}$. 
The chemical potential $\mu$ can then be calculated for given 
$J$, $U_{12}$, and $\bar{n}$. 
In addition, the problem is characterized by the wave vectors for the 
flow in each of the components, $(k_1, k_2)$. In Sec.\ \ref{sec:model}, 
we derived the detailed form of the Bogoliubov equations for 
one moving condensate, $(k_1,k_2)=(0,-k)$; and opposite currents, 
$(k_1,k_2)=(k/2,-k/2)$.

\subsection{Weak coupling and broken translational symmetry}
\label{sec:weak}

\begin{figure}
\includegraphics[width=0.95\columnwidth]{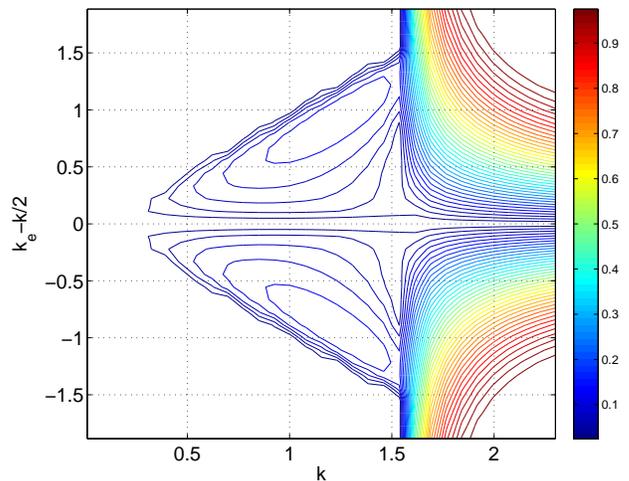} 
\caption{(Color online) 
Growth rate of the dynamically most unstable mode 
(i.e. the imaginary part of the mode with the largest imaginary frequency) 
as a function of flow wavevector $k$ and and mode wavevector $k_e$ (note that
a shift $k_e-k/2$ has been applied to the y-axis to make the symmetry
around $k_e=k/2$ more transparent). 
Here, one component is in motion with wavevector $k$ and the other is 
stationary. The lattice has $256$ sites, the 
parameter values are $J/U=1$, $U_{12}/U=1.1$, and 
the asymptotic density is $\bar{n}=1$. 
}
\label{fig:bogweak}
\end{figure}

We first consider the case $J/U=1$, $U_{12}/U = 1.1$, $\bar{n}=1$. 
Figure \ref{fig:bogweak} shows the stability properties of the system 
in the case of one moving condensate, $(k_1,k_2)=(0,-k)$. 
The figure plots the largest imaginary excitation frequency as a function 
of flow wavevector $k$ and excitation wavevector $k_e$; thus, a 
vertical cross-section in the figure at a fixed $k$ gives the imaginary 
part of the spectrum for a given current. 
The lattice is of finite size, $256$ sites, in the numerics (in fact, real-life lattices may be of similar dimensions) 
and this influences the numerical value of the imaginary part somewhat. The effect is such that
the imaginary part becomes larger as the lattice size is reduced to $128$.
The physics at low wavevector $k$ 
is clearly recognized from the case of a pair of classical 
fluids or a homogeneous two-component BEC 
\cite{takeuchi2010,suzuki2010,chandrasekhar1961book,gerwin1968RMP}.
For a given $k$, not too large, 
a range of excitations with wavevectors $0 < |k_e-k/2| \lesssim k$ 
become unstable.
The amplitudes of the unstable modes are localized at the interface, indicating 
that these are indeed interface modes. 

We compare to the case of a continuous, i.e., non-lattice system. 
In Ref.\ \cite{suzuki2010}, it was observed that in the case of 
a narrow interface, the binary condensate displays the classical 
dispersion relation 
\cite{chandrasekhar1961book}
\beq
\omega = \sqrt{\frac{\alpha}{2\rho}k_e^3 - v^2k^2} \quad {\rm (classically),}
\label{chandrasekhar}
\enq 
where $v$ is the velocity of the unperturbed flow and $\alpha$ is 
the surface tension. However, if the interface is wider, the upper 
stability line lies close to the line $|k_e-k/2|=k$; this was in 
Ref.\ \cite{suzuki2010} termed the counter-superflow instability.
Apparently, Fig.\ \ref{fig:bogweak} is in this latter regime, and 
the discreteness is not seen to cause considerable deviations from the linear 
stability line. 

We note a few curious features in Fig.\ \ref{fig:bogweak}. 
For very small $k$, the instability seems to be absent; 
this will be discussed further in Sec.\ \ref{sec:strong}. 
For wavevectors exceeding a critical value, $k>k_{c}=\pi/2$, 
it is seen in Fig.\ \ref{fig:bogweak} that all excitation 
wavevectors become unstable. This is no longer an interface  
mode, but an instability known to occur also in single-component 
discrete BECs  \cite{desarlo2005}. 
The physics behind the instability is that the single-particle dispersion 
relation, $\epsilon_k = J(1-\cos k)$, has an inflection point at 
$k_c$, and thus the effective mass goes from positive, to diverging, 
to negative. This is well understood and studied in single-component 
BECs and we will not dwell on it further, but only conclude that the 
interesting range of $k$ values for the case $(k_1,k_2)=(0,-k)$ 
ranges between $-\pi/2$ and $\pi/2$.

\begin{figure}
\includegraphics[width=0.95\columnwidth]{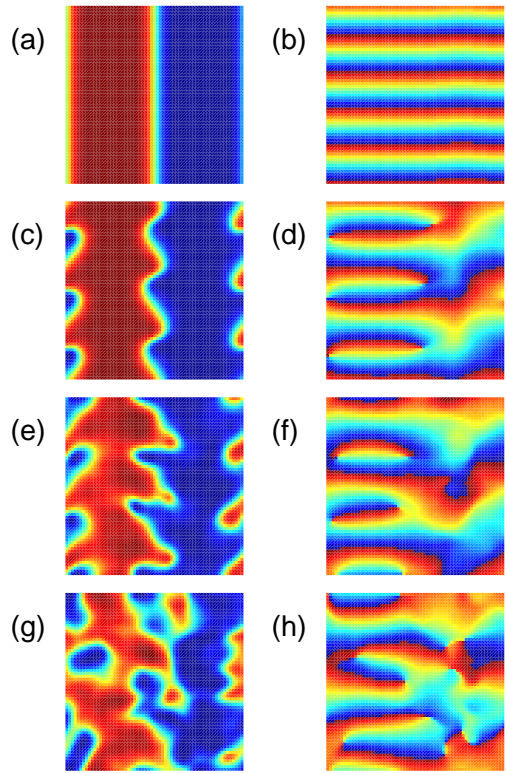} 
\caption{(Color online) 
Time development of the instability for the case $k=5\pi/64$, and 
remaining parameter values as in Fig.\ \ref{fig:bogweak}. 
Panels (a),(c),(e),(g) show the density of the 1 component, where 
red indicates high density and blue is zero density; panels 
(b), (d), (f), (h) show the corresponding phase. Snapshots are taken at 
times $t=44$ for (a)-(b); $t=176$ for (c)-(d); $t=265$ for (e)-(f); and 
$t=400$ for (g)-(h). 
}
\label{fig:timeweak}
\end{figure}

To illustrate the long time behavior of
Kelvin-Helmholtz instability, we simulate the time 
development of the discrete GP equation. A series of snapshots are shown 
in Fig.\ \ref{fig:timeweak} for the case $k=5\pi/64\approx 0.245$. 
The time development is simulated using a split-step Fourier technique, 
with a grid of size 64$\times$64 sites and periodic boundary conditions 
along both directions. For the chosen 
wavevector $k=5\pi/64$, a surface wave with a dominant wavevector 
$k_e=3\pi/64$ can be seen to
grow exponentially until the nonlinear stage is reached, 
where the condensates break up into smaller fragments.

\begin{figure}
\includegraphics[width=0.95\columnwidth]{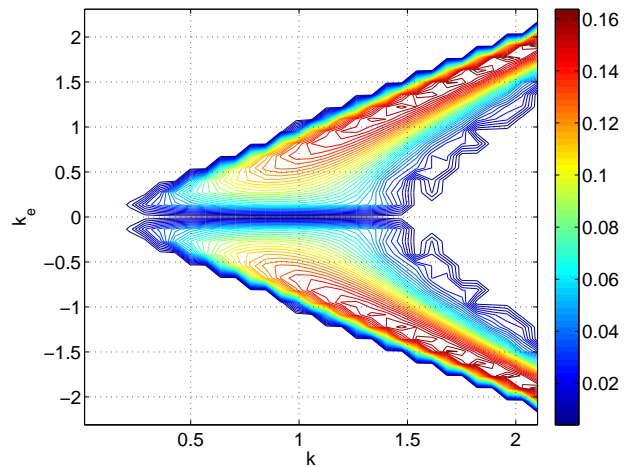} 
\caption{(Color online) 
Largest imaginary part of the excitation frequency as in 
Fig.\ \ref{fig:bogweak}, but with counter-flowing condensates 
having wavevectors $(k/2,-k/2)$. The lattice has $256$ sites, 
parameter values are $J/U=1$, $U_{12}/U=1.1$, and 
the asymptotic density is $\bar{n}=1$. 
}
\label{fig:weaksymmetric}
\end{figure}

For the case of counterflow, $(k/2,-k/2)$, 
the result for the imaginary parts of the modes is shown in Fig.\ 
\ref{fig:weaksymmetric}. 
By definition of $k$, the critical relative 
wavevector for negative effective mass is in this case $k_c=\pi$. 
For large enough $k$, it is seen that the modes with 
small wavevectors $k_e$ are actually stabilized. 
We thus define a second critical wavenumber for the flow, $k_s$, above 
which long-wavelength excitations are stabilized. 
The corresponding flow velocity is denoted $v_s$. 
From Figure \ref{fig:weaksymmetric}, we read off $k_s \approx 1.5$ 
(the closeness to $\pi/2$ is suggestive but, in fact, 
a coincidence). 
This is a counterpart to the classically 
known stabilization of long wavelengths when the flow velocity 
exceeds $\sqrt{8}$ times the sound speed $c$ \cite{gerwin1968RMP}, 
\beq
v_s = \sqrt{8}c \quad {\rm (classically).} 
\enq 
However, in the discrete case, there 
seems to be no simple relationship between the onset of 
long-wavelength stability and sound speed. 
In this particular case, the flow velocity in each component 
is $v_s=2J\sin(k_s/2)\approx 1.4U$ in dimensionless units, whereas the 
sound speed is $c=\sqrt{JU\bar{n}} = U$, 
but further numerical experimentation 
shows only a weak dependence of $v_s$ on $c$; this puzzle will be resolved 
in Sec.\ \ref{sec:analytic}.

\begin{figure}
\includegraphics[width=0.45\columnwidth]{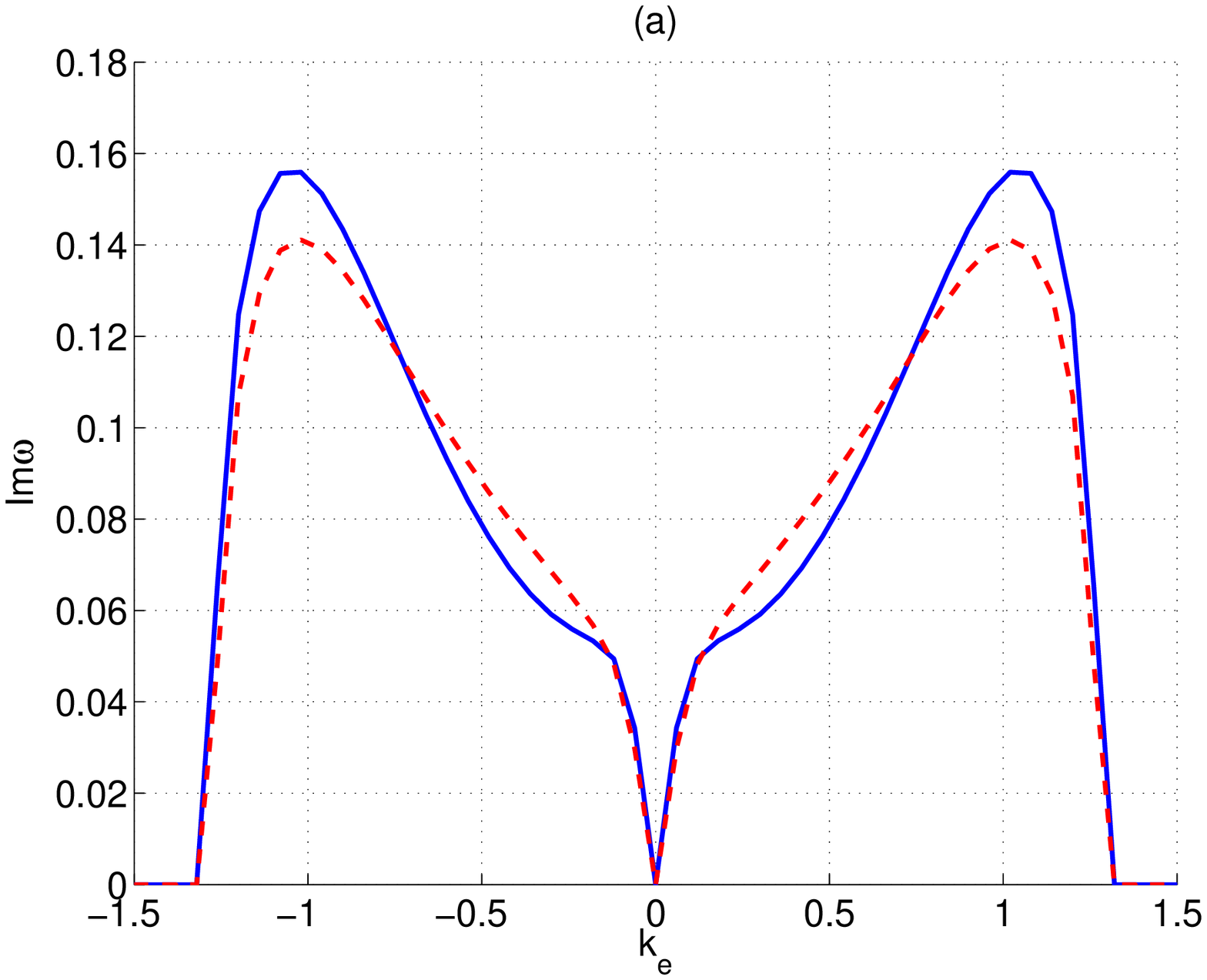} 
\includegraphics[width=0.45\columnwidth]{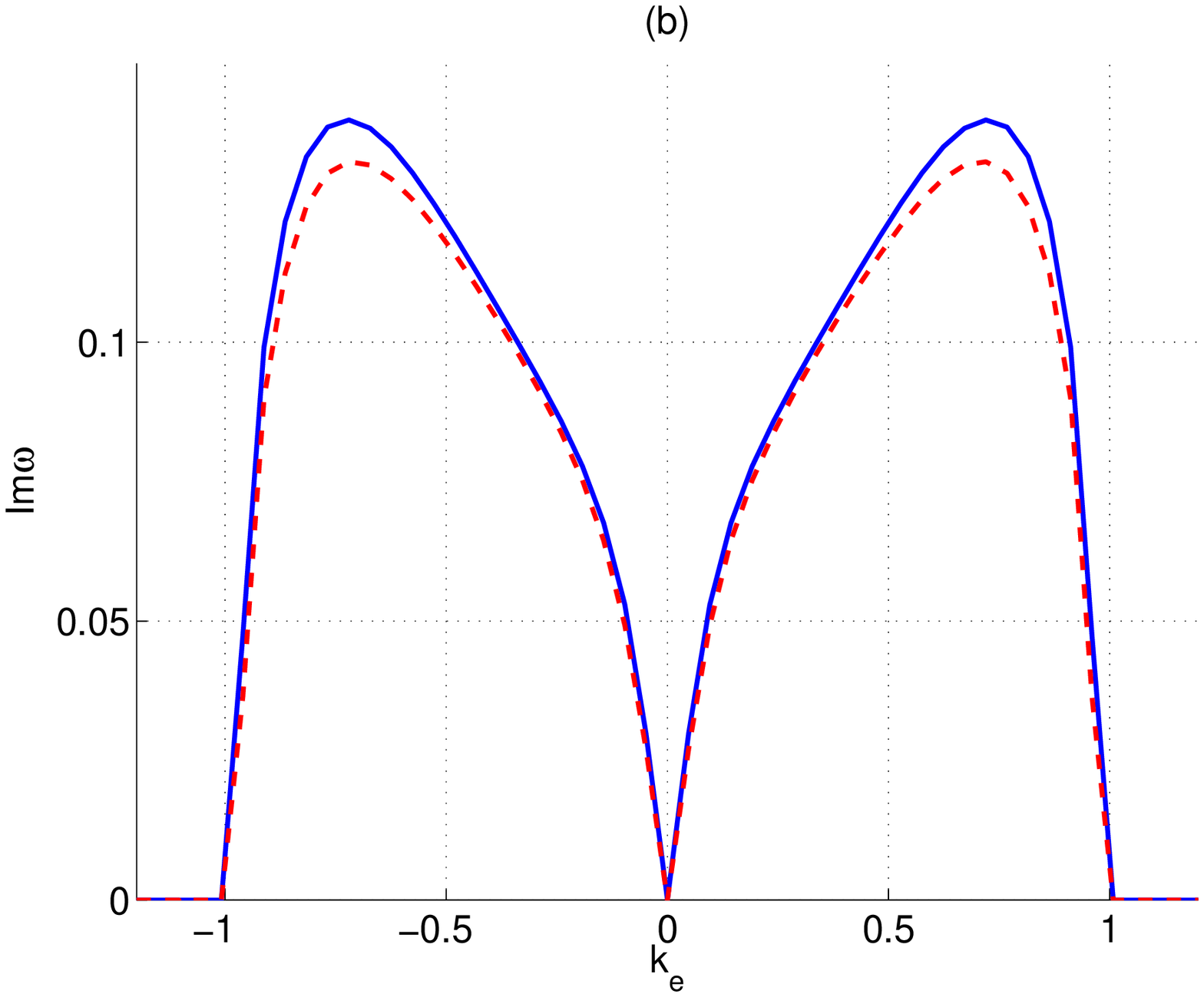} 
\caption{(Color online) Comparison of the largest imaginary part of the excitation frequency 
as a function of the excitation wavenumber $k_e$, in the cases 
of one flowing component, $(k,0)$ (dashed red), and counterflow, 
$(k/2,-k/2)$ (solid blue). 
Parameter values are as in Fig.\ \ref{fig:bogweak}, but 
the wavevector is fixed at $k=1.3$ in (a), and $k=1.0$ in (b). 
For the case of one flowing 
component, the $k_e$ axis is shifted by $k_e\to k_e-k/2$. 
The lattice has $256$ sites, 
parameter values are $J/U=1$, $U_{12}/U=1.1$, and 
the asymptotic density is $\bar{n}=1$. 
}
\label{fig:compare_k_k2}
\end{figure}

Since the system lacks Galilean invariance, the spectrum for a given 
flow wavevector $k$ 
is not invariant under a transformation $(k_1,k_1) \to (k_1-q,k_2+q)$. 
We plot as an example in Fig.\ \ref{fig:compare_k_k2} the imaginary 
frequency for  the same parameters as above, for two choices of $k$,
comparing the cases $(k,0)$ and $(k/2,-k/2)$. 
It is seen that the curves 
are similar, but the symmetric case $(k/2,-k/2)$ is more unstable, in 
the sense that it has a higher maximum imaginary frequency. Further 
numerical experimentation (not shown here) indicates that this 
observation holds quite generally.

\subsection{Narrow interface physics: Effects of discreteness} 
\label{sec:strong}

\begin{figure}
\includegraphics[width=0.95\columnwidth]{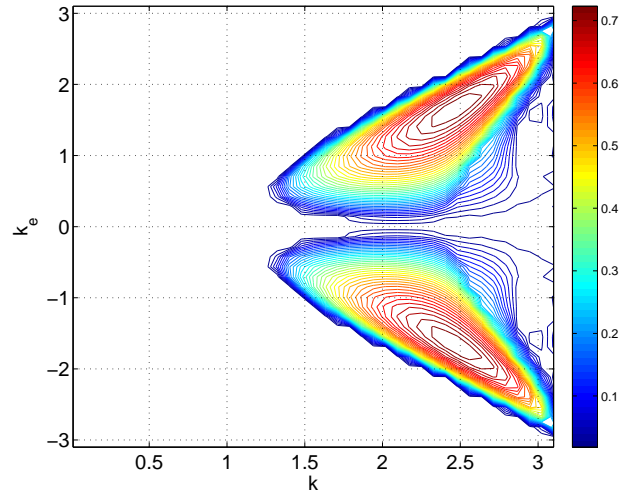} 
\caption{(Color online) 
Growth rate of the dynamically most unstable mode 
(i.e. the imaginary part of the mode with the largest imaginary frequency) 
as a function of flow wavevector $k$ and and mode wavevector $k_e$. 
Here, both components are in motion with counter-directed wavevectors 
$(k/2,-k/2)$.
The parameters are $J/U=1$, $U_{12}/U=1.5$, and 
the asymptotic density is $\bar{n}=4$. 
}
\label{fig:bogstrong}
\end{figure}

Figure \ref{fig:bogstrong} shows the stability spectrum of a system 
at a higher density, $\bar{n}=4$, and a stronger repulsion between the 
components, $U_{12}=1.5U$. 
The combined effect of increased density, which means larger 
interaction energy, and increased inter-species repulsion, 
implies a narrower interface \cite{ao1998}. 
Already in the case $\bar{n}=1, U_{12}=1.1U$, we saw that low flow velocities 
appeared to be stable, and here the effect is more pronounced. 
This feature of the instability is not seen in simulations of 
homogeneous BECs \cite{takeuchi2010,suzuki2010}, nor in classical 
fluids, in the absence of external forces \cite{chandrasekhar1961book}. 
We attribute this to the discrete density profile at 
the interface, as will be detailed below. 

In the figure, we note also that there is a stable region for 
large flow wavevectors $k>k_s$, just as in the case $\bar{n}=1$, 
but now $k_s\approx 2.8$. 
We will return to this in 
Sec.\ \ref{sec:analytic}. Finally, we observe that the 
upper instability line at large $k$ values is still 
approximately linear with unit slope, 
indicating that we are in the counter-superflow regime despite the 
narrower interface. 

We now turn to the stability of slow flow wavevectors $k$. 
First recall how a dynamical instability in the 
Bogoliubov equations result from a collision between a quasiparticle mode 
and a quasihole mode \cite{lundh2006b}. Starting from a reference point 
in phase space -- here, take $k=0$ -- we consider a quasiparticle 
mode ${\bf v}_a$ and a quasihole mode ${\bf v}_b$ with real, nearly 
degenerate eigenfrequencies $\omega_{0a}$ and $\omega_{0b}$. 
We then slightly increase a control parameter -- in this case, 
$k$ -- and thus transform the Bogoliubov matrix $M\to M'=M+\delta M$. 
Writing ${\bf v}=a{\bf v}_a+b{\bf v}_b$ and solving the Bogoliubov equations 
for the new eigenfrequencies $\omega$, we find in the general case
\begin{widetext}
\beq
\omega = \frac{\omega_{0a}+\omega_{0b}+\delta\omega_a+\delta\omega_b}{2}
\pm \sqrt{\left(\frac{\omega_{0a}-\omega_{0b}+\delta\omega_a-\delta\omega_b}{2}
\right)^2 
- |X|^2},
\enq
\end{widetext}
where $\delta\omega_a={\bf v}_a^{\dag}\delta M{\bf v}_a$, 
$\delta\omega_b=-{\bf v}_b^{\dag}\delta M{\bf v}_b$, and 
$X={\bf v}_a^{\dag}\delta M{\bf v}_b$. In other words, when the two modes 
become degenerate, the cross-term makes them unstable. 

In the present case, we consider two modes related by 
the symmetry of Eq.\ (\ref{symmetry}); thus, we have 
$\omega_{0a}+\delta\omega_a=-\omega_{0b}-\delta\omega_b$, and 
\beq
\omega = \pm \sqrt{\left(\omega_{0a}+\delta\omega_a\right)^2 - |X|^2}.
\enq
In a homogeneous, i.e., non-discrete system, the lowest-lying 
surface mode corresponds to a uniform translation of the boundary 
and is thus a Goldstone mode; at $k=k_e=0$ it has zero energy, 
i.e., $\omega_{0a}=0$ in the limit $k_e=0$. 
With increasing $k_e$, its energy increases from zero.
A finite $k$ results in a shift in the Bogoliubov matrix $\delta M$ as 
discussed above and the Kelvin-Helmholtz instability comes about 
because $|X|$ is always greater than $\delta \omega_a$ in these 
systems. The onset of instability therefore tends towards 
$k=0$ as $k_e$ tends to zero. 

In a discrete system, however, translational symmetry is lost and 
there is no Goldstone mode: A small shift of the boundary 
profile costs a finite amount of energy. This translates into 
$\omega_{0a}=\epsilon>0$ at $k_e=0$, which in turn creates a threshold value 
of $k$ for the onset of instability. 
For stronger interactions or higher densities, the interface is 
thinner, and the effects of discreteness are more pronounced. 
This explains the increased threshold value for higher density. 

\begin{figure}
\includegraphics[width=0.95\columnwidth]{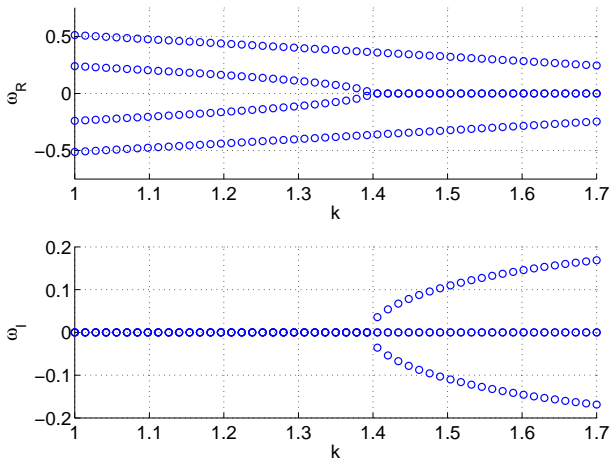} 
\caption{
Few Bogoliubov eigenvalues for a fixed excitation wavevector $k_e=0.3$, 
as functions of the flow wavevector $k$. Parameters are as in 
Fig.\ \ref{fig:bogstrong}. Upper panel: Real parts, lower panel: 
Imaginary parts. 
}
\label{fig:spectrumfork}
\end{figure}

\begin{figure}
\includegraphics[width=0.95\columnwidth]{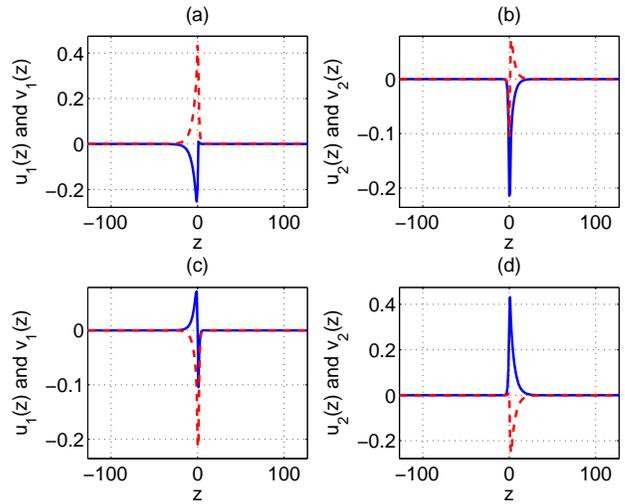}
\caption{Bogoliubov eigenstates for 
the lowest-lying surface modes,
functions $u_j(z)$ (blue solid lines) and $v_j(z)$ 
(red dashed) at $k_e=0.3$ and 
$k=1.25$, which is below the instability threshold. 
Upper panels: Quasiparticle mode; Lower panels: Quasihole mode. 
Left panels: mode functions $u_1(z), v_1(z)$ associated with component 1, 
and right panels: mode functions $u_2(z), v_2(z)$.
The lattice has $256$ sites, 
parameter values are $J/U=1$, $U_{12}/U=1.5$, and 
the asymptotic density is $\bar{n}=4$.
}
\label{fig:modeprofiles}
\end{figure}

Figure \ref{fig:spectrumfork} displays some of 
the lowest-lying Bogoliubov eigenvalues 
for a fixed $k_e=0.3$, as functions of $k$, at $\bar{n}=4$. 
It is clearly seen how instabilities result from a merging of 
two eigenstates with equal but opposite energies, thus related by 
the symmetry in Eq.\ (\ref{symmetry}). For large $k$ similar instabilities
happen for many other modes and here, for clarity, we plot just the appearance 
of the first instability.
The modes responsible for the 
dominant instability start out at a small but finite energy for 
$k=0$; this energy does not tend to zero as $k_e\to 0$ 
(although the latter fact cannot be inferred from the figure). 
The mode functions $u_i, v_i$ are plotted in Fig.\ \ref{fig:modeprofiles}. 
These are the lowest-lying surface modes corresponding to translation 
of the surface; notice how the symmetry (\ref{symmetry}) is 
manifest in the plots.

\section{Towards an analytical description}
\label{sec:analytic}

In order to obtain a feeling for the mathematics and physics behind 
the Kelvin-Helmholtz instability, we show here a heuristic derivation 
of an approximate analytical formula for the excitation frequencies. 
In effect, we replace the spatially dependent coefficients in 
Eq.\ (\ref{bdg_1d}) by c-numbers. This can be justified by expanding 
the solution $\mathbf{v}$ in a suitable basis \cite{nilsen2008}. 
It can also be effected by simply assuming a local-density approximation 
at the center of the interface.
We discuss the more symmetric case with counter-directed currents, 
$(k_1,k_2)=(k/2,-k/2)$, since it allows for solutions on closed form. 
The matrix is approximated as 
\beq
 {\hat M}=\left(\begin{array}{cccc}
\epsilon+\Delta_++g  & g & C & C\\
g & \epsilon + \Delta_- + g & C & C\\
C & C & \epsilon + \Delta_- + g & g\\
C & C & g & \epsilon + \Delta_+ + g
\end{array}\right),\nonumber
\label{constantmatrix}
\enq
where $g$ is proportional to the in-species interaction energy, and $C$ is a measure of the inter-species 
repulsion. $\epsilon$ is a self-energy associated with bending 
in the $z$ direction of the respective mode function, which is to be 
identified with the finite energy $\omega_{0a}$ of the almost-Goldstone modes 
dicussed above. 
Finally, $\Delta_{\pm}$ are the relative kinetic energies. 
In the discrete lattice case, 
\beq
\Delta_{\pm} = -2J(1-\cos k/2)+2J\left[1 - \cos(k_e \pm k/2)\right],
\label{variationalkin1}
\enq
and in the continuous case without a lattice, 
\beq
\Delta_{\pm} = \frac{\hbar^2}{2m} (k_e^2 \pm k_e k).
\label{variationalkin2}
\enq 
The continuous case is simplest: Inserting the dispersion, putting 
$\hbar=m=1$ for convenience, we obtain 
the eigenfrequencies
\begin{widetext}
\bea
\omega^2 = (2g+ \frac12 k_e^2 + \epsilon)(\epsilon+\frac12 k_e^2) + \frac14 k_e^2 k^2 
\pm 2\sqrt{\epsilon + \frac12 k_e^2}
\sqrt{(2g + \epsilon+\frac12 k_e^2)\frac14 k_e^2k^2 + C^2(\epsilon+\frac12 k_e^2)
}.
\label{analytic_dispersion}
\ena
\end{widetext}
This equation is of the same form as that in Ref.\ \cite{suzuki2010}, describing the counter-superflow instability. 
Clearly, the case with a minus sign lets $\omega^2$ become negative for 
certain parameter values, and thus describes the unstable modes. 
The discrete case yields somewhat clumsier expressions, but can be solved; 
we discuss the differences below. 

The choice of parameters can be done with the benefit of hindsight; 
however, assuming a local-density approximation in the midpoint of the 
interface, where in the limit of a broad interface 
the densities $n_1$ and $n_2$ are half their bulk values, we readily obtain 
$g=U{\bar{n}}/2=\mu/2$, and $C=U_{12}\bar{n}/2=(U_{12}/U)g$. 
The self-energy $\epsilon$ is smaller than the other energies, 
and so is $C-g$, so 
we first put $\epsilon=0$ and $C=g$ in Eq.\ (\ref{analytic_dispersion}) and 
find the points where $\omega^2$ turns negative; these are the boundaries 
for instability. We obtain the solutions 
\bea
k_e = 0, \nonumber\\
k = k_e, \nonumber\\
k^2-k_e^2 - 4\mu = 0.
\ena 
Linearizing around the first instability, we obtain 
\beq
\omega^2 = \frac12 k_e^2 \left( \epsilon-\frac12 k^2 \right) + 
{\mathcal O}(k_e^4),
\enq
which is the well-known linearly increasing imaginary frequency, existing 
for $k^2 < 4\mu$. The modes are stabilized again when $k_e>k$. 
The upper stability limit lies at $k_e=k$ as observed in Ref.\ \cite{suzuki2010}. 
However, when $k^2 > 4\mu$, small wavevectors $k_e$ are stabilized. 
This is similar to the well known classical stabilization at 
relative flow velocities exceeding $\sqrt{8}$ times the 
sound velocity, as we discussed in Sec.\ \ref{sec:weak} \cite{gerwin1968RMP}. 
However, even in the continuous case, the prefactor does not match with 
the classical result (nor with Ref.\ \cite{suzuki2010}); we simply 
attribute this to the quite different kind of dispersion relation. 

A small but finite $\epsilon$ will introduce small corrections to the 
above. In order to salvage the linear dispersion at small excitation 
wavevectors $k_e$ -- which is observed to be the case numerically -- we must put 
$C=g+\epsilon/2$. 
With a finite $\epsilon$, slow flow velocities are stabilized. This happens 
only in the discrete case, but in order to bring out the qualitative 
features while displaying reasonably compact formulas,
we show here the bastardized expression resulting from using a 
continuous-case dispersion relation together with a finite self-energy. 
Expanding the eigenfrequency in powers of $k$ and $k_e$, one obtains 
\beq
\omega^2 = \frac12k_e^2 (\frac12 k_e^2 + \epsilon) 
-  \frac14 k^2 k_e^2 + {\mathcal O}(k^6, k_e^6, k^2 k_e^4, k^4 k_e^2),
\enq
which is positive for $k^2 < q^2+2\epsilon$. 
Thus, slow flow velocities are seen to be absolutely stable. Classically, 
this only happens in the presence of gravity or an equivalent asymmetric 
force \cite{chandrasekhar1961book,takeuchi2010}, but in the discrete 
system, there is stabilization even in the absence of forces.

The discrete-case dispersion relation leads to more complicated algebra, 
but all the qualitative conclusions remain the same. 
When $\epsilon=0$ and $g=C$, the frequency is again zero for $k_e=0$ and 
$k_e=k$; a finite $\epsilon$ stabilizes small $k$. 
The most important modification is that of the sound-speed 
limit, where we now obtain stability at small $k_e$ if 
$k>k_s$, with $k_s$ given by 
\beq
2J\cos(\frac{k_s}{2})=-\mu/2 + \sqrt{\mu^2/4+4J^2},
\enq
whose solution is 
always real. The continuous-case expression is recovered when 
$\mu/J$ is small; then the flow velocity is 
$v_s = 2J\sin k_s \approx \sqrt{4\mu}$, similar to the continuous case. 

Furthermore, for $k > \pi/2$, the spectrum changes character completely 
and a wide range of excitation wavevectors $k_e$ become unstable;
this is again a manifestation of the bulk instability 
associated with negative effective mass of 
Ref.\ \cite{desarlo2005}.

Thus, a heuristic constant-amplitude calculation has managed to 
bring out all the features present in the numerical results. 
The familiar linear increase of imaginary frequency at small $k_e$ for 
a given flow wavevector $k$; the stabilization at excitation 
wavevectors $k_e$ exceeding $k$; the absolute stabilization of 
slow relative velocities; the stabilization of small $k_e$ for 
$k$ above a cutoff depending on the sound velocity; and even the 
bulk instability when the effective mass becomes negative, 
could all be described.

\section{Strongly correlated regime}
\label{sec:mott}
The discussion so far has been limited to the case of two Bose gases 
in the Bose-Einstein condensed state. This is the state of the system 
when the ratio $J/U$ is not very small and when the density is not very 
small. In the general case, one must account for quantum fluctuations 
whose most striking effect is to drive a quantum phase transition 
between the superfluid and Mott insulating phases \cite{jaksch1998}. 
In addition, quantum fluctuations are seen to shift the transition line 
between mixed and phase-segregated phases, as we shall see below.

\begin{figure}
\includegraphics[width=0.95\columnwidth]{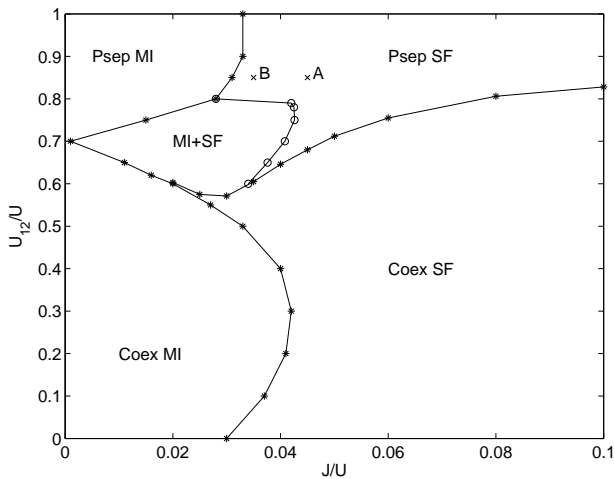} 
\caption{Phase diagram for a mixture of two identical Bose 
gases, calculated in the Gutzwiller approximation. 
The chemical potential is taken to be $\mu=0.7U$. 
The points A and B mark the parameters for which the 
two simulations described in the text are done. 
}
\label{fig:phasediagram}
\end{figure}

The many-body Hamiltonian (\ref{hamiltonian}) has been seen to display a 
rich phase diagram \cite{powell2009,iskin2010}. In Figure \ref{fig:phasediagram}, 
we display the portion of greatest interest for our purposes. 
The phase diagram is calculated in the Gutzwiller approximation, 
which is explained below. 
The general picture is that for weak enough hopping $J$, except in 
degenerate cases, both gases are in the Mott insulating state 
(marked MI in Fig.\ (\ref{fig:phasediagram})) with 
suppressed on-site number fluctuations and zero superfluid density; 
conversely, for large enough $J$ the bosons are always superfluid (SF). 
If the inter-component interaction $U_{12}$ is large enough 
compared with the chemical potential $\mu$, then the two components 
are phase separated (marked Psep), either Mott insulating or superfluid; 
and if $U_{12}$ is small enough or $\mu$ big enough, the two 
components coexist (marked Coex). 
There exists one further phase close to the line $U_{12}=\mu$ at 
weak hopping, where one component is Mott insulating and the other 
one is superfluid with a lower density (marked SF+MI in the phase 
diagram). 
Finally, note that for a 
pair of Bose-Einstein condensed gases, the transition between 
phase separated and coexisting phases takes place at 
$U_{12}=U$. In contrast, in the limit of weak $J$, 
phase separation takes place at $U_{12}=\mu$.

The Gutzwiller approximation is based on a mean-field ansatz for 
the many-body state, 
\beq
|\psi_G(t)\rangle = \prod_r \prod_{j=1}^{2}  |\phi_{r,j}(t)\rangle.
\label{gutzwiller} 
\enq
This ansatz amounts to treating both the hopping between sites and the 
inter-species interaction in a mean-field fashion. 
This can be done since on-site entanglement between the two species 
is not expected to be important. For computations, it is convenient 
to expand the on-site states in a local Fock basis with an upper 
cutoff $n_{\rm max}$, 
\beq
|\phi_{r,j}(t)\rangle = \sum_{n=0}^{n_{\rm max}} C_{r,j,n}(t) |n\rangle_{r,j}. 
\enq
In the Gutzwiller approximation, the local total density $n$ and condensate 
wave function $z$ for each of the components is readily computed. We 
characterize the system by studying the behavior of local density $n_j$, the 
local condensate density $n_{cj}=|\Phi_j|^2$, 
and phase $\varphi_j={\rm arg} \Phi_j$.
In order to calculate the ground state, the Hamiltonian 
(\ref{hamiltonian}) is minimized with respect to the complex coefficients 
$C_{r,j,n}$. To simulate dynamics, we propagate the coupled equations 
of motion \cite{zakrzewski2005,lundh2011}
\beq
i \frac{\partial C_{r,j,n}}{\partial t} = 
\frac{\delta}{\delta C_{r,j,n}^*} \langle \psi_G(t)| H |\psi_G(t)\rangle. 
\enq
For the present simulation of Kelvin-Helmholtz physics, the initial 
condition was constructed by calculating the phase separated ground 
state, then imprinting counter-directed currents on the two components, 
and finally adding a small amount of random noise onto each of the 
complex coefficients $C_{r,j,n}$.

We simulate interface dynamics in the ``Psep SF'' phase. 
In the Mott insulating phase, dynamics is quite trivially absent. 
In the ``SF+MI'' phase, there is also no interface dynamics: 
Since the Mott-superfluid phase transition is second order, there 
is no surface tension and hence no capillary waves are expected 
to form between superfluid and Mott insulating regions. 
Phenomena connected to melting of Mott insulators are 
conceivable at high enough energies (cf.\ \cite{lundh2011,krutitsky2011}), but we will 
not pursue this here.

\begin{figure}
\includegraphics[width=0.95\columnwidth]{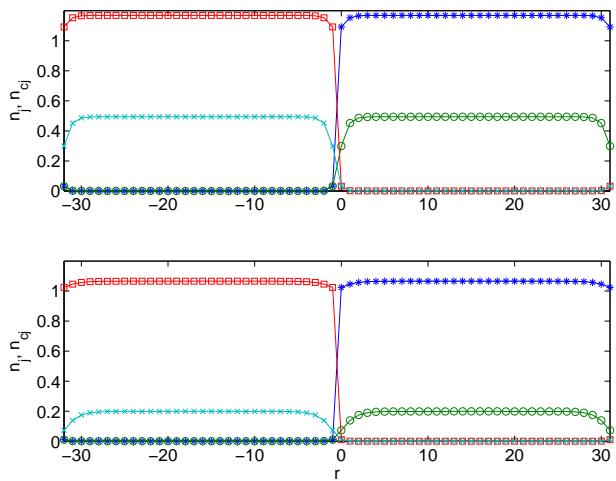} 
\caption{(Color online) Cross sections of the density profile of two phase 
segregated Bose gases at points A (upper panel) and B (lower panel) 
in the phase diagram, respectively. Squares denote the density for 
component 1 $n_1$; crosses denote $n_{c1}$; asterisks $n_2$, 
and circles $n_{c2}$. Lines are to guide the eye. 
}
\label{fig:crossectionmott}
\end{figure}

\begin{figure}
\includegraphics[width=0.95\columnwidth]{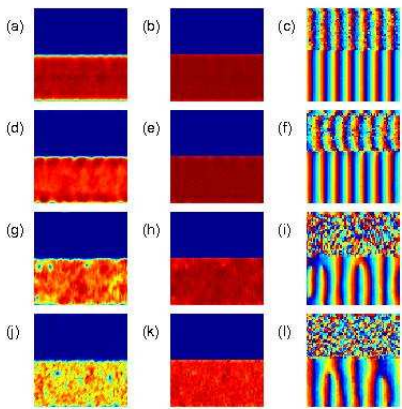} 
\caption{(Color online)
Time development in the Gutzwiller approximation 
of a system with $J=0.045U$, $U_{12}=0.85U$, and 
$\mu=0.7U$, with wave vectors $(k/2,-k/2)$ where $k=7\pi/16$. 
Snapshots displayed in the different rows are taken at times 
(from top to bottom) $100/U$, $200/U$, $300/U$, and $4000/U$. 
Left column indicates total density $n$, middle column 
condensate density $n_c$, and right column the phase $\varphi$, 
of bosons of component 1. The corresponding quantities for 
component 2 are not shown, but are similar to the ones shown 
mirrored in the $x$ axis. 
}
\label{fig:mottdynamics1}
\end{figure}

The first simulation is done at the point marked A in the phase 
diagram of Fig\ \ref{fig:phasediagram}: 
$J=0.045U, U_{12}=0.85U$, and $\mu=0.7U$. Here, we have two 
phase separated superfluids, and as seen in Fig.\ \ref{fig:crossectionmott}, 
the Bose-Einstein condensed part 
of the fluid is less than half the total density. 
Also the superfluid density is expected 
to be small close to the Mott transition \cite{rey2003}. 
Figure \ref{fig:mottdynamics1} shows the dynamics of a system with 
symmetrically imprinted wavevectors $(k/2,-k/2)$, where 
$k=7\pi/16\approx 1.3744$. 
It is seen that the condensate densities behave very much like 
in the pure-condensate case. The time scale is vastly different 
due to the difference in tunneling matrix element $J$. In addition, 
it is seen that the total density is affected very little, 
and on the time scales we have simulated the system seems to be 
settling down into a phase separated steady state. 
The two fluids do not mix; 
the main effect of the KH instability is to excite vortices within the 
respective components. 
In the phase plots one can see vortices as phase singularities; 
they show up as zeros of the condensate density but the total density 
is not depleted. These are vortices filled with Mott insulating 
atoms, as previously described in Refs.\ \cite{wu2004,lundh2008b}. 

\begin{figure}
\includegraphics[width=0.95\columnwidth]{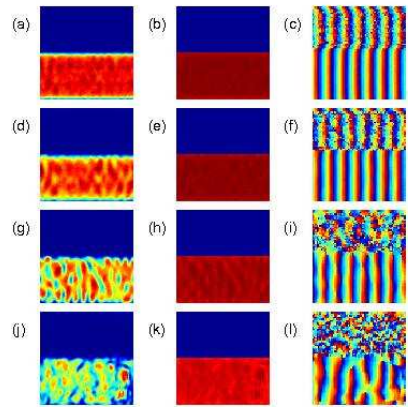} 
\caption{(Color online)
Time development in the Gutzwiller approximation 
of a system with $J=0.035U$, $U_{12}=0.85U$, and 
$\mu=0.7U$, with wave vectors $(k/2,-k/2)$ where $k=7\pi/16$. 
Snapshots displayed in the different rows are taken at times 
(from top to bottom) $100/U$, $200/U$, $300/U$, and $4000/U$. 
Left column indicates total density $n$, middle column 
condensate density $n_c$, and right column the phase $\varphi$, 
of bosons of component 1. The corresponding quantities for 
component 2 are not shown, but are similar to the ones shown 
mirrored in the $x$ axis. 
}
\label{fig:mottdynamics2}
\end{figure}

We next show a simulation at the point marked B in the phase 
diagram of Fig\ \ref{fig:phasediagram}: 
$J=0.035U, U_{12}=0.85U$, and $\mu=0.7U$. This is still in the 
regime of two phase separated superfluids, but even closer to the Mott 
transition. 
Figure \ref{fig:mottdynamics2} shows the dynamics.
We see that the dynamics is in fact more rapid here than in the 
case of larger $J$. 
At times $t=200/U$ and $t=300/U$, a significant wave pattern develops 
in the condensate component, and at $t=1000/U$, the system contains 
many more vortices than in the case in Fig.\ \ref{fig:mottdynamics1}. 
The total density is less affected, but exhibits some protruding 
features at the surface. 
Clearly, with a smaller condensate 
fraction, the threshold energy is smaller for modulation of 
condensate density and formation of vortices.
Close to the transition, where the condensate density decreases 
rapidly towards zero, this overcomes the effect of the modest 
reduction of $J$.

\section{Conclusions}
\label{sec:conclusion}
In summary, we have studied the Kelvin-Helmholtz instability on the 
interface between two lattice Bose gases in relative motion. 
The instability is seen to be affected by three effects introduced by 
the lattice potential: broken translational symmetry, 
discreteness, and quantum fluctuations.  
Broken translational symmetry affects the instability in such a way 
that the excitation frequencies do not depend only on the relative velocity, 
but on the flow velocities of the two gases separately; the symmetric 
case of two counter-flowing gases is seen to be the most unstable situation. 
Second, the discreteness will stabilize slow relative currents, so that 
the instability is prevented if the relative velocity is low enough. 
Finally, strongly correlated physics affects the physics in such a way 
that a neighboring Mott insulating phase will prevent the two Bose 
systems from mixing after the Kelvin-Helmholtz instability is first excited. 
Close to the Mott phase transition, only the superfluid density will 
take part in the instability, but the total density will hardly be affected. 

\begin{acknowledgments}
E.L. acknowledges support from the Swedish Research Council (VR). 
J.-P. M acknowledges support from the Academy of Finland (Project 135646).
Authors also acknowledge useful discussions with Prof. Vitaly Bychkov.
Calculations have been conducted using the resources of 
High Performance Computing Center North (HPC2N).
This project was initiated during 
the NORDITA Program ``Quantum solids, liquids and gases''. 
\end{acknowledgments}


\end{document}